\documentstyle[twocolumn,aps,epsfig]{revtex}

\begin{document}

\wideabs{

\title{Surface modes and critical velocity in trapped Bose-condensates.}

\author{A.E. Muryshev\protect\( ^{1}\protect \) and P.O. Fedichev\protect\( ^{1,2}\protect \)}

\address{\protect\( ^{1)}\protect \) Russian Research Center ``Kurchatov Institute'',
Kurchatov Square, 123182, Moscow, Russia\\
\protect\( ^{2)}\protect \)Institute f. Theoretische Physik, Universitat Innsbruck,
Technikerstr. 25/2, A6020, Innsbruck, Austria}

\maketitle
\begin{abstract}
We propose the novel mechanism leading to superfluidity breakdown in dilute
Bose-condensed gases. We discuss the properties and highlight the role played
by surface excitations in trapped condensates. We show that the critical velocity
measured in recent experiments is nothing else but the critical velocity associated
with the spontaneous creation of surface modes. The latter turns out to be larger
than that found by the analysis of a superflow stability with respect to
vortex nucleation. 
\end{abstract}
}

Superfluidity is one of the most striking manifestations of quantum statistics
in many-body interacting systems. It is usually defined as ``the property of
a liquid to move without friction inside capillaries or around obstacles''.
Superfluidity is confined to small flow velocities. There is a certain critical
value of the fluid speed \( v_{c} \) beyond which the dissipation settles in.
The value of the critical velocity is directly related to excitations properties
\cite{LL:volIX,feynman:smbook}. The analysis of the energy conservation in
a moving liquid shows \cite{LL:volIX} that the flow becomes thermodynamically
\emph{}unstable when the liquid velocity exceeds
\begin{equation}
\label{Landaucrit}
v_{c}=\min _{p}\frac{\epsilon (p)}{p}.
\end{equation}
 \emph{}Here \( \epsilon (p) \) is the energy of an excitation with the momentum
\( p \) in the laboratory frame. 

When applied to liquid helium, the criterion (\ref{Landaucrit}) gives the critical
velocity coinciding with the sound velocity, which is too high to explain the
experimentally measured value. Moreover, the experimental results appear to
be non-universal and depend on the geometry of the liquid sample \cite{Donelly}.
This contradiction was resolved by R. Feynman \cite{feynman:smbook,Donelly}
who suggested, that the dissipation is related to creation of quantized vortices.
If this is taken into account, the critical velocity turns out to be both sufficiently
small and depends on the external factors. The combination of the vortex hypothesis
and the Landau criterion (\ref{Landaucrit}) very well explains the experiments
in liquid helium \cite{Donelly}. 

These fundamental theories are now being tested in the new experiments with
ultra-cold Bose-condensed gases (see \cite{Ketterle:review,Pitaevskii:review}
for a review of experimental and theoretical activity in the field of Bose-Einstein
condensation (BEC) in weakly interacting gases). It is now possible to study
dynamical properties of superfluidity in systems, which are both exceptionally
well controlled experimentally and understood theoretically. The closest analogy
for the critical velocity measurement in long capillaries is the experiment
done at MIT \cite{Ketterle:critvel}. There the onset of the dissipation caused
by a moving laser beam was studied as a function of the beam velocity. The reported
value of the critical velocity is well below the speed of sound. This can be
explained either by the vortex nucleation \cite{Ketterle:critvel,Zwerger:critvel},
or by the modification of the excitation spectrum in inhomogeneous condensates
\cite{fedichev:critvel}.

Another way to measure critical velocity is to rotate a superfluid and find
the minimum (critical) angular velocity \( \Omega _{c} \) at which an excitation
shows up in the system. The straightforward generalization of Eq.(\ref{Landaucrit})
for the case of a rotating superfluid gives \cite{feynman:smbook,LL:volIX}:
\begin{equation}
\label{Landau:rot}
\Omega _{c}=\min _{M}\frac{\epsilon (M)}{M},
\end{equation}
where \( \epsilon (M) \) is the energy of an excitation as a function of its
angular momentum \( M \). In a Bose-condensed gas, the lowest value of \( \Omega _{c} \)
is obtained when \( \epsilon (M) \) is the energy of a single vortex state.
Remarkably, the critical velocity (\ref{Landau:rot}) is smaller than that observed
at ENS \cite{dalibard:critvel}.

In this Letter we perform a simple analysis of critical velocity issues in a
trapped Bose-condensed gas. We point out that the universal criterion (\ref{Landaucrit})
gives no information about the dissipation time scales. Actually, Eqs. (\ref{Landaucrit})
and (\ref{Landau:rot}) merely reflect the fact that dissipation is not forbidden
by the energy conservation law as soon as the (angular) velocity of the liquid
reaches a certain value. In spite of the fact that vortex excitations may have
the lowest values of the critical velocity, in a dilute gas they can hardly
be formed suddenly out of nothing. Instead, as it is prompted both by the experiment
\cite{dalibard:quadr} and numerical simulations \cite{ueda:surfacenum}, vortices
are nucleated at the condensate border from the condensate shape oscillations.
This can only happen if the velocity of the superflow exceeds the critical velocity
corresponding to the formation of the necessary surface modes, which later evolve
into the vortices in the course of non-linear condensate dynamics. To support
this view we analyze the elementary excitations of a confined Bose-condensed
gas. We show that, in a sufficiently dense condensate, the lowest possible values
of the critical velocity correspond to the excitations localized at the low
density region close to the edge of the condensate. Remarkably, the surface
modes turn out to be responsible for the previously reported decrease of \( v_{c} \)
compared to the velocity of sound in inhomogeneous condensates \cite{fedichev:critvel}.
We note that this effect is a novel feature characteristic of dilute Bose-condensed
gases and is absent in ``conventional'' superfluids, such as liquid helium.
In the latter case the superfluid is incompressible and the surface excitations
of the mentioned type cannot exist. Using a simple model of the surface modes,
we calculate both \( v_{c} \) (for an MIT-type experiment) and \( \Omega _{c} \)
(for an ENS-type experiment), assuming that the generation of the surface modes
is the necessary precondition for the vortex nucleation. 

Consider a Bose-condensate confined in an infinitely long cylindrical harmonic
trap characterized by the transverse frequency \( \omega  \) (hereafter we
assume \( \hbar =m=1 \), where \( m \) is the mass of the condensate atom).
The condensate is characterized by its chemical potential \( \mu  \) and is
assumed to be in the Thomas-Fermi(TF) regime: \( \mu \gg \omega  \). In this
case, the condensate density has a simple form: \( n_{0}(\rho )=n_{0}(1-\rho ^{2}/R_{c}^{2}), \)
where \( \rho  \) is the radial coordinate and \( R_{c}=(2\mu /\omega ^{2})^{1/2} \)
is the TF size of the condensate. The condensate density on the axis of the
trap \( n_{0} \) is related to the chemical potential: \( \mu =4\pi n_{0}a \),
where \( a \) is the scattering length \cite{Pitaevskii:review}.

The character of the excitations in this geometry is pretty well understood
by now. The lowest elementary excitations are phonons. The velocity of sound
is found to be \( \sqrt{2} \) times smaller than that in a homogeneous condensate
of the same density \( n_{0} \): \( c_{Z}=\sqrt{gn_{0}/2} \) \cite{Zaremba:soundprop}.
The radial profile of the phonon modes turns out to be very similar to the profile
of the condensate wavefunction. Hence, the phonons are delocalized inside the
condensate and are analogous to the ``bulk'' phonons in homogeneous Bose-condensates
or in liquid helium. As the axial wave-vector of the excitations \( p \) increases
(i.e. when \( pR_{c}\sim 1 \)), the radial dependence of the phonon modes becomes
more and more involved, and the spectrum ceases to be linear \cite{Zaremba:soundprop}.
The behavior of the excitations for arbitrary values of \( p \) was studied
numerically in \cite{fedichev:critvel}. The critical velocity obtained with
the help of Eq.(\ref{Landaucrit}) was shown to decrease together with the TF
parameter \( \eta =\omega /\mu  \) and can be much smaller than the velocity
of sound \( c_{Z} \).

As we will see below, the excitations with the lowest critical velocity are
in fact the surface modes, i.e. the excitations localized at the TF border of
the condensate. In this region the condensate wavefunction \( \psi _{0} \)
has a universal form and can be analyzed with the help of the following universal
equation \cite{pitaevskii:condensateborder,pethick:condensateborder}:
\begin{equation}
\label{gpborder}
-\frac{1}{2}\Delta \chi _{0}+\chi _{0}^{3}+x\chi _{0}=0.
\end{equation}
Here \( x=(\rho -R_{c})/d \), \( d=R_{c}(\omega ^{2}/2\mu ^{2})^{1/3}\ll R_{c} \)
is the ``condensate border thickness'', and \( \chi _{0}=(d^{2}g)^{1/2}\psi _{0} \)
is the dimensionless condensate wavefunction. Deeply inside the condensate spatial
region ( \( x\rightarrow -\infty  \)), the solution \( \chi _{0} \) matches
the TF density profile: \( \chi _{0}\rightarrow \sqrt{-x} \). Far away from
the condensate border (\( x\rightarrow \infty  \) ) the condensate density
quickly decreases: \( \chi _{0}\rightarrow 0.19\exp (-2x^{3/2}/3)/x^{1/2} \)
\cite{pitaevskii:condensateborder}.

\begin{figure}
{\par\centering \resizebox*{0.9\columnwidth}{!}{\includegraphics{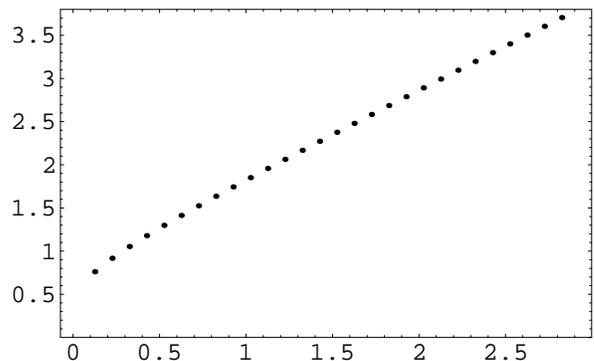}} \par}

\caption{Dimensionless spectral function \protect\( F\protect \) for the lowest surface
excitations obtained by numerical solution of Eqs.(\ref{fminuni}) and (\ref{fpluuni}).\label{funcF}}
\end{figure}

The spectrum of elementary excitations can be found by studying the behavior
of quantized linear fluctuations of the order parameter \( \delta \hat{\psi }=\hat{\psi }-\psi _{0} \),
where \( \hat{\psi } \) is the annihilation operator of the atomic field \cite{LL:volIX},
and
\[
\delta \hat{\psi }=\sum _{pm}(\hat{a}_{pm}u_{pm}e^{im\phi +ipz}-\hat{a}^{\dagger }_{pm}v_{pm}e^{-im\phi -ipz}).\]
Here \( \hat{a}_{pm} \),(\( \hat{a}^{\dagger }_{pm} \)) are the annihilation(creation)
operators of the Bogolyubov excitations characterized by the longitudinal and
angular momentums \( p \) and \( m \), respectively. One can show that the
functions \( f_{pm}^{\pm }=u_{pm}\mp v_{pm} \) satisfy the following Bogolyubov-de
Gennes equations \cite{Ohberg}:
\begin{equation}
\label{fmin}
\epsilon _{pm}f_{pm}^{-}=(-\frac{\Delta _{\rho }}{2}+\frac{m^{2}}{2\rho ^{2}}+\frac{p^{2}}{2}+\frac{\Delta \psi _{0}}{2\psi _{0}})f^{+}_{pm},
\end{equation}
\begin{equation}
\label{fplu}
\epsilon _{pm}f_{pm}^{+}=(-\frac{\Delta _{\rho }}{2}+\frac{m^{2}}{2\rho ^{2}}+\frac{p^{2}}{2}+\frac{\Delta \psi _{0}}{2\psi _{0}}+2g\psi _{0}^{2})f^{-}_{pm}.
\end{equation}
 The whole spectrum of the elementary excitations can be recovered by solving
Eqs.(\ref{fmin}),(\ref{fplu}) numerically together with the exact Gross-Pitaevskii
equation for the condensate wave function \( \psi _{0} \). Below we will take
advantage of the representation (\ref{gpborder}) and analyze only the modes
confined to the vicinity of the condensate border \( |\rho -R_{c}|\sim d \).
This can be done by rewriting Eqs.(\ref{fmin}),(\ref{fplu}) in the same units
as Eq.(\ref{gpborder}). Since the excitations are radially localized inside
a shell with the thickness \( d\ll R_{c} \), the Laplacian \( \Delta  \) can
be substituted by the second derivative in \( x \), and the centrifugal term
can be approximated by its value at \( \rho =R_{c} \). Thus, Eqs.(\ref{fmin}),(\ref{fplu})
take the simple universal form
\begin{equation}
\label{fminuni}
\tilde{\epsilon }_{pm}f_{pm}^{-}=(-\frac{d^{2}}{2dx^{2}}+\frac{k^{2}}{2}+\frac{\Delta \chi _{0}}{2\chi _{0}})f^{+}_{pm},
\end{equation}
\begin{equation}
\label{fpluuni}
\tilde{\epsilon }_{pm}f_{pm}^{+}=(-\frac{d^{2}}{2dx^{2}}+\frac{k^{2}}{2}+\frac{\Delta \chi _{0}}{2\chi _{0}}+2\chi _{0}^{2})f^{-}_{pm},
\end{equation}
where \( \tilde{\epsilon }=\epsilon d^{2} \), and 
\[
k^{2}=(p^{2}d^{2}+\frac{m^{2}d^{2}}{R_{c}^{2}}).\]
Since Eqs.(\ref{fminuni}),(\ref{fpluuni}) contain only the single parameter
\( k^{2} \), the spectrum of the excitations is given by the universal relation:
\( \epsilon _{pm}=d^{-2}F(k^{2}) \). Here \( F \) is a dimensionless spectral
function found numerically (see Fig. \ref{funcF}).

To make sense of the numerical results we resort to a simple model, based on
WKB approximation to Eqs.(\ref{fminuni}) and (\ref{fpluuni}). Quasiclassical
radial momentum \( p_{\rho } \) has the following Bogolyubov-type form
\begin{equation}
\label{prho}
\frac{p_{\rho }^{2}}{2}+\frac{k^{2}}{2}=\sqrt{\tilde{\epsilon }_{pm}^{2}+\chi _{0}^{4}}-\chi _{0}^{2}-\frac{\chi _{0}^{\prime \prime }}{2\chi _{0}}.
\end{equation}
To make the things even more simple, we ignore the exact behavior of the condensate
wave function close to the condensate border \( |x|\alt 1 \) and substitute
\( \chi _{0} \) with its asymptotic values at \( |x|\gg 1 \). For sufficiently
high momentums (\( k^{2}\gg 1 \)), we find, that the classical turning points
are close to the condensate border \( x=0 \). The dimensionless energy of the
excitations is much larger than the mean field interaction \( \chi _{0}^{2} \).
Hence, the dispersion relation (\ref{prho}) can be written as
\[
\frac{p_{\rho }^{2}}{2}+\frac{k^{2}}{2}=\tilde{\epsilon }_{pm}+|x|.\]
 Born-Sommerfeld quantization rule gives essentially the single-particle spectrum
\begin{equation}
\label{surface:spectrum}
\epsilon _{pm}(n)=\frac{C_{n}}{d^{2}}+\frac{p^{2}}{2}+\frac{m^{2}}{2R_{c}^{2}},
\end{equation}
where \( n=0,1,... \) is the radial quantum number, and \( C_{n}=(3\pi /8\sqrt{2})^{2/3}(4n+1)^{2/3} \)
is the dimensionless constant found from the boundary condition at the classical
turning points \cite{LL:volIII,migdal:kach}. The quasiclassical value \( C_{0}=0.885 \)
found in our model calculation is remarkably close to that obtained by exact
numerical solution of Eqs.(\ref{fminuni}) and (\ref{fpluuni}): \( C_{0}\approx 0.9 \).
This coincidence is somewhat surprising, since our analysis fails to recognize
the complicated character of the condensate wavefunction close to the condensate
border. At the same time the numerical solution shows that the mode functions
are localized at \( |x|\agt 1 \), where our approximations are better justified.
Moreover, WKB approximation is known to provide very reasonable estimations
for excitation energies even if the quantum numbers are small \cite{migdal:kach}.
The exact solution of Eqs. (\ref{fminuni}) and (\ref{fpluuni}) also shows
that the approximation (\ref{surface:spectrum}) holds fairly well even for
\( pd,md/R_{c}\sim 1 \), in spite of the fact that the excitations characterized
by small momentums \( k\alt 1 \) spread more inside the condensate spatial
region and, hence, are not single particles (see Fig. \ref{funcF}).

We note that, although the surface modes (\ref{surface:spectrum}) arise from
the analysis of the Bogolyubov dispersion relation, they are quite different
from the Bogolyubov excitations in a homogeneous condensate. At smaller momentums
(\( pd\alt 1) \) the localization size of the solutions to Eqs. (\ref{fminuni}),(\ref{fpluuni})
increases and gradually approaches the size of the condensate. In this limit,
the dispersion relation recovers its phonon limit. The surface modes (\ref{surface:spectrum})
exist in a thin layer around \( \rho =R_{c} \) and which is still thicker than
the healing length \( l_{0}=1/\sqrt{8\pi an_{0}} \) in the condensate: \( d\gg l_{0} \)
in the TF limit. This may happen only in a compressible liquid and thus the
dispersion relation similar to (\ref{surface:spectrum}) can hardly be found
in ``traditional'' superfluids, such as liquid helium. In the latter case
the condensate density drops on a distance scale of order \( l_{0} \) at the
wall of the confining vessel.

The validity of Eq.(\ref{surface:spectrum}) is restricted by a number of assumptions.
First, we require that the radial localization length of the surface excitations
be much smaller than the size of the condensate. Since we are mostly interested
in \( \epsilon _{pm}\sim 1/d^{2}, \) the radial size of the modes is \( \sim d\ll R_{c} \)
in the TF regime. In the same way we can neglect the spatial variations of the
centrifugal potential provided that the momentums of the involved excitations
are sufficiently small. Indeed, a simple WKB calculation shows that the first
correction to Eq. (\ref{surface:spectrum}) arising from the variations of the
centrifugal potential close to \( \rho \approx R_{c} \) results in a slight
change of the ground state energy: \( \Delta C_{0}/C_{0}\sim m^{2}\mu /R_{c}^{2} \).
The solution of Eq.(\ref{Landau:rot}) corresponds to \( m\sim R_{c}/d\ll R_{c}/\mu ^{1/2} \),
so that for our purposes, the finite-\( m \) corrections to \( C_{n} \) can
be safely neglected. 

The dispersion relation (\ref{surface:spectrum}) is the central result of our
work. It allows us to study the critical velocities (\ref{Landaucrit}) and
(\ref{Landau:rot}) in the deep TF regime (\( \eta \ll 1 \)). First we turn
to an MIT-type experiment. The application of criterion (\ref{Landaucrit})
shows that the lowest value of critical velocity corresponds to the instability
of the excitations with \( m=0 \): 
\begin{equation}
\label{vc}
v_{c}=\sqrt{2C_{0}/d^{2}}\sim 1/d\sim c_{Z}\eta ^{1/3}\ll c_{Z}.
\end{equation}
 This result has to be compared with the stability analysis of the vortex excitations.
Consider a superfluid flow of velocity \( v_{s} \) without vortices. The energy
of the liquid scales as \( \sim \rho _{s}v_{s}^{2}R_{c}^{2}L \), where \( \rho _{s}\approx n_{0} \)
is the superfluid density, and \( L \) is the length of the trap. Let us consider
instead the vortex configuration similar to that discussed in relation to the
experiment \cite{ketterle:phasesing}, i.e. a sequence of vortices at the distance
\( \pi /v_{s} \) from each other along the trap axis. The average velocity
of the superfluid would be the same \( v_{s} \). The energy of this vortex
chain is roughly the sum of the vortex energies and thus can be estimated as
\( \rho _{s}R_{c}\log (1/\eta )v_{s}L \). Comparing the energy of the superflow
with and without the vortices, we find that the vortex configuration has a lower
energy as soon as \( v_{s}>v_{c}^{(v)}\sim \eta \log (\eta ^{-1})c_{Z} \).
This means that in a TF condensate, vortices form the true ground state already
at \( v>v_{c}^{(v)} \). At the same time, since the vortices are nucleated
only at the edge of the condensate, they originate from the surface excitations,
which only become unstable if \( v>v_{c}>v_{c}^{(v)} \). Therefore, in the
TF regime no vortices can be nucleated unless the superfluid velocity reaches
\( v_{c} \). This statement does not contradict the conclusions derived in
\cite{ketterle:phasesing}, since the vortex states posses very small energies
compared to the irrotational superflow of velocity \( v_{c} \), and therefore
further relaxation leads to the creation of vortices. 

The dispersion relation for the surface excitations also gives us a way to discuss
the critical angular velocity in an ENS-type rotating superfluid experiment.
Again, applying the criterion (\ref{Landau:rot}) to the dispersion relation
(\ref{surface:spectrum}) we immediately find that the surface modes in a cylindrical
trap first become unstable if \( p=0 \) and 
\begin{equation}
\label{omegac}
\Omega >\Omega _{c}=\frac{v_{c}}{R_{c}}\sim \omega \eta ^{1/3}.
\end{equation}
This value has to be compared with the lowest angular velocity at which a vortex
can be formed in the condensate. It can be derived using the same kind of simple
arguments as above: \( \Omega _{c}^{(v)}\sim \omega \eta \log (\eta ^{-1}) \)
\cite{LL:volIX}. As in the discussion above, in the deep TF limit \( \Omega _{c}^{(v)}<\Omega _{c} \)
and, we conclude that, in order to make up a vortex on the surface of a rotating
superfluid, one has to first reach the angular velocity (\ref{omegac}). Only
then the surface modes are nucleated and, finally, a vortex appears in the superfluid
in the course of the turbulent liquid dynamics. This point of view can be supported
by earlier work \cite{dalibard:quadr}, where the authors conclude that \( \Omega _{c} \)
coincides with the frequencies of the phonon-like excitations with \( m=2,3 \)
(depending on the symmetry of the rotating perturbation). We argue that the
TF parameter in the ENS experiment is not sufficiently small. Since the critical
velocities (\ref{vc}) and (\ref{omegac}) change very slowly as \( \eta  \)
decreases, the distinction between the bulk and the surface modes is not very
sharp at moderate values of the TF parameter. 

Finally, we would like to comment on the physical nature of criteria (\ref{Landaucrit})
and (\ref{Landau:rot}). Consider an obstacle (probe) of a size \( l \) moving
inside a superfluid with the velocity \( v \). In a weakly interacting Bose-condensed
gas the dissipation arises pair creation of excitations with radial momentums
\( p \) and \( p^{\prime } \). The energy conservation law \cite{LL:volIX}
requires that
\begin{equation}
\label{enconsrv}
\epsilon _{p}+\epsilon _{p^{\prime }}-v(p+p^{\prime })=0.
\end{equation}
If the fluid velocity slightly exceeds \( v_{c} \) (\ref{Landaucrit}), both
solutions of Eq.(\ref{enconsrv}) \( p,p^{\prime }\approx p_{*} \), where \( p_{*} \)
is found from the equation \( \epsilon (p_{*})=v_{c}p_{*} \). At the same time,
the uncertainty principle requires \( (p+p^{\prime })\alt 1/l \). Therefore,
\( p_{*}l\alt 1 \) and thus the dissipation at \( v\approx v_{c} \) is only
possible if the probe is sufficiently small: \( l\alt p^{-1}_{*} \). In our
case we find that \( p_{*}\sim d^{-1} \) and thus the probe dissipates only
if \( l\alt d \). In this sense the criteria (\ref{Landaucrit}) and (\ref{Landau:rot})
give the values of the critical velocities measured by an ideal, point size
probe. We note that this requirement is not at all automatically satisfied in
the experiments. This, in turn, may lead to additional complications in ascribing
a particular mechanism to the reported critical velocity values. For example,
a finite size probe may nucleate a vortex ring on its surface. This instability
leads to a sufficiently small value of the critical velocity: \( v_{c}\sim 1/l \)
(see \cite{Zwerger:critvel} and refs. therein). In the latter case, the mechanism
of the critical velocity decrease is somewhat similar to that reported here:
the increase of the superfluid velocity around a moving body leads to the decrease
of the condensate density, so that the moving body-superfluid interface can
host an unstable ``surface'' excitation.

In conclusion, we develop a simple theory of critical velocity in trapped BEC.
We highlight the role played by the surface excitations in large TF condensates.
We suggest that, since the vortex excitations are formed at the condensate border
and thus are initially surface excitations, the critical velocity measured in
the recent experiments cannot be smaller than the critical velocity associated
with the generation of the surface excitations. The latter turns out to be larger
than that found by the analysis of thermodynamic stability of superflow with
respect to vortex nucleation both in a long trap with axially moving superfluid
and in a rotating trapped Bose-condensate. We note that this novel feature is
a direct consequence of compressibility of trapped Bose-condensed gases and
hence cannot be related to traditional superfluids, such as liquid helium. 

We acknowledge fruitful discussions with G.V. Shlyapnikov and W. Zwerger. The
present work was supported by Austrian Science Foundation, by INTAS, and Russian
Foundation for Basic Research(RFBR, grant 99-02-18024).

\bibliographystyle{prsty}
\bibliography{myDb}

\begin{thebibliography}{10}

\bibitem{LL:volIX}
E.~M. Lifshitz and L.~P. Pitaevskii, {\em Statistical Physics, Part 2}
  (Pergamon Press, Oxford, 1980).

\bibitem{feynman:smbook}
R.~P. Feynman, {\em Statistical Mechanics: a set of lectures} (Benjamin
  Reading, Massachusetts, 1982).

\bibitem{Donelly}
R.~J. Donnelly, {\em Quantized vortices in helium II} (Cambridge University
  Press, Cambridge, 1991).

\bibitem{Ketterle:review}
W. Ketterle, D.~S. Durfee, and D.~M. Stamper-Kurn, {\em Proceedings of the
  International School of Physics "Enrico Fermi"}, edited by M.Inguscio, S.
  Stringari, and C. Wieman (IOS press, Amsterdam, 1999).

\bibitem{Pitaevskii:review}
F. Dalfovo, S. Giorgini, L. Pitaevskii, and S. Stringari, Rev. Mod. Phys. {\bf
  71},  463  (1999).

\bibitem{Ketterle:critvel}
C. Raman {\it et~al.}, Phys. Rev. Lett. {\bf 83},  2502  (1999).

\bibitem{Zwerger:critvel}
J.~S. Stiessberger and W. Zwerger, 2000, cond-mat/0006419.

\bibitem{fedichev:critvel}
P.~O. Fedichev and G.~V. Shlyapnikov, Phys. Rev. A {\bf 63},  045601  (2001).

\bibitem{dalibard:critvel}
F. Chevy, K.~W. Madison, and J. Dalibard, Phys. Rev. Lett {\bf 85},  2223
  (2000).

\bibitem{dalibard:quadr}
F. Chevy, K.~W. Madison, V. Bretin, and J. Dalibard, cond-mat/0104218.

\bibitem{ueda:surfacenum}
M. Tsubota, K. Kosamatsu, and M. Ueda, cond-mat/0104523.

\bibitem{Zaremba:soundprop}
E. Zaremba, Phys. Rev. A {\bf 57},  518  (1998).

\bibitem{pitaevskii:condensateborder}
F. Dalfovo, L.~P. Pitaevskii, and S. Stringari, Phys. Rev. A {\bf 54},  4213
  (1996).

\bibitem{pethick:condensateborder}
E. Lundh, C.~J. Pethick, and H. Smith, Phys. Rev. A {\bf 55},  2126  (1997).

\bibitem{Ohberg}
P. \"{O}hberg {\it et~al.}, Phys. Rev. A {\bf 56},  R3346  (1997).

\bibitem{LL:volIII}
L.~D. Landau and E.~M. Lifshitz, {\em Quantum Mechanics, Non-relativistic
  Theory} (Pergamon Press, Oxford, 1977).

\bibitem{migdal:kach}
A.~B. Migdal, {\em Qualitative methods in quantum theory} (W. A. Benjamin, INC,
  London, 1977).

\bibitem{ketterle:phasesing}
S. Inouye {\it et~al.}, 2001, cond-mat/0104444.

\end{thebibliography}

\end{document}